\documentclass[twocolumn,showpacs,amsfonts,aps,prl,nofootinbib,floatfix,%
superscriptaddress]{revtex4-1}

\usepackage{amsmath}
\usepackage{bm}
\usepackage{graphicx}

\newcommand{\dNdy}{dN_\mathrm{ch}/dy}

\newcommand{\be}[1]{\begin{equation}\label{#1}}
\newcommand{\ee}{\end{equation}}

\newcommand{\eq}{{\,=\,}}

\def\La{\langle}
\def\Ra{\rangle}

\usepackage{color}

\begin{document}


\title{$\bm{200\,A}$\,GeV Au+Au collisions serve a nearly perfect
       quark-gluon liquid}


\author{Huichao Song}
\affiliation{Nuclear Science Division,
             Lawrence Berkeley National Laboratory, Berkeley,
             California 94720, USA}
\affiliation{Department of Physics, The Ohio State University,
             Columbus, Ohio 43210, USA}

\author{Steffen A. Bass}
\affiliation{Department of Physics, Duke University, Durham,
             North Carolina 27708, USA}

\author{Ulrich Heinz}
\affiliation{Department of Physics, The Ohio State University,
             Columbus, Ohio 43210, USA}

\author{Tetsufumi Hirano}
\affiliation{Department of Physics, The University of Tokyo, Tokyo 113-0033,
             Japan}
\affiliation{Nuclear Science Division,
             Lawrence Berkeley National Laboratory, Berkeley,
             California 94720, USA}

\author{Chun Shen}
\affiliation{Department of Physics, The Ohio State University,
             Columbus, Ohio 43210, USA}

\begin{abstract}
A new robust method to extract the specific shear viscosity
$(\eta/s)_\mathrm{QGP}$ of a Quark-Gluon-Plasma (QGP) at temperatures
$T_\mathrm{c}{\,<\,}T{\,\alt\,}2T_\mathrm{c}$ from the centrality
dependence of the eccentricity-scaled elliptic flow $v_2/\varepsilon$
measured in ultra-relativistic heavy-ion collisions is presented.
Coupling viscous fluid dynamics for the QGP with a microscopic transport
model for hadronic freeze-out we find for $200\,A$\,GeV Au+Au
collisions that $v_2/\varepsilon$ is a universal function of
multiplicity density $(1/S)(\dNdy)$ that depends only on the viscosity
but not on the model used for computing the initial fireball
eccentricity $\varepsilon$. Comparing with measurements we
find $1<4\pi(\eta/s)_\mathrm{QGP}<2.5$ where the uncertainty range is
dominated by model uncertainties for the values of $\varepsilon$ used to
normalize the measured $v_2$.
\end{abstract}
\date{\today}
\maketitle


Ever since heavy-ion collision experiments at the Relativistic Heavy-Ion
Collider (RHIC) demonstrated the creation of color-deconfined
Quark-Gluon Plasma (QGP) exhibiting almost ideal fluid dynamical collective
behavior \cite{Heinz:2001xi,Gyulassy:2004zy,Arsene:2004fa}, with viscosity
per entropy density $\eta/s$ approaching the KSS lower bound
$\frac{\eta}{s}{\,\agt\,}\frac{1}{4\pi}$
\cite{Policastro:2001yc,Danielewicz:1984ww}, an accurate extraction of
the QGP transport coefficients, especially its shear viscosity
$(\eta/s)_\mathrm{QGP}$, from experimental measurements has been of great
interest \cite{Molnar:2001ux,Romatschke:2007mq,Song:2007fn}.
A small $\eta/s$ is generally considered to be evidence for the onset
of a strongly-coupled deconfined plasma early in the evolution of the
collision. RHIC is the first accelerator to provide sufficient beam energy
for the QGP to live long enough for flow observables to become sensitive to
its intrinsic transport properties. Simulations based on both viscous fluid
dynamics and quark-gluon transport theory \cite{Molnar:2001ux,%
Romatschke:2007mq,Song:2007fn} have established
that the elliptic flow generated in non-central heavy-ion collisions is
particularly sensitive to the shear viscosity $\eta/s$ of the medium. However,
a quantitative extraction of $(\eta/s)_\mathrm{QGP}$ from elliptic flow data
requires not only accurate elliptic flow measurements but also a precise
knowledge of the theoretical baseline corresponding to zero QGP viscosity.
The latter, in turn, requires good control over the fluid's collective
response to anisotropic pressure gradients, and a realistic microscopic
description of chemical and kinetic freeze-out during the hadronic
stage \cite{Song:2008hj}. A purely hydrodynamic approach that treats
both the dense early QGP and dilute late hadron resonance gas phases
as viscous fluids not only requires the introduction of two additional
parameters, the chemical and kinetic freeze-out temperatures, which must
be separately adjusted to experimental data, but ultimately fails
\cite{Song:2010aq} because viscous corrections due to hadronic dissipation
are large \cite{Hirano:2005xf} and invalidate a fluid dynamical approach
even if it properly accounts for chemical decoupling before kinetic
freeze-out \cite{Teaney:2001av,Huovinen:2009yb} and for a strong growth
\cite{Demir:2008tr} of the specific shear viscosity $\eta/s$ in the
hadronic stage \cite{Shen:2011kn}.

We here use a newly developed hybrid code (see \cite{Song:2010aq} for
details) that couples the relativistic (2+1)-dimensional viscous fluid
algorithm {\tt VISH2{+}1} \cite{Song:2007fn} to the microscopic hadronic
scattering cascade {\tt UrQMD} \cite{Bass:1998ca} via a Monte Carlo
interface \cite{fn-1}. For the QGP fluid we assume constant $\eta/s$ for
$T_\mathrm{c}{\,<\,}T{\,\alt\,}2T_\mathrm{c}$ \cite{Csernai:2006zz}.
We switch from a hydrodynamic description of the QGP to {\tt UrQMD} at
temperature $T_\mathrm{sw}\eq165$\,MeV, adjusted to reproduce the chemical
freeze-out temperature measured in RHIC collisions \cite{BraunMunzinger:2001ip}
and the highest $T$ for which we have a valid microscopic description.
By giving us full microscopic control, without additional parameters, over
the complex hadron kinetic freeze-out our hybrid model opens the door for
quantitatively exploring the transport properties of the earlier QGP phase
using measured final hadron spectra.

For the hydrodynamic evolution above $T_\mathrm{sw}$ we use the
state-of-the-art equation of state (EOS) {\tt s95p-PCE}
based on recent lattice QCD results \cite{Huovinen:2009yb}. The remaining
model uncertainties arise mainly from the initial conditions of the
hydrodynamic evolution, including the starting time $\tau_0$ and
initial transverse flow velocity. While these cannot be directly
measured and require model input, they are tightly constrained by
experimental information on the final state \cite{Heinz:2001xi,Hirano:2007ei}.
Modeling the QGP as an ideal fluid with $\eta/s\eq0$ and zero initial
transverse flow requires an early start at $\tau_0\eq0.4$\,fm/$c$.
Non-zero shear viscosity adds to the transverse pressure
\cite{Molnar:2001ux,Romatschke:2007mq,Song:2007fn}, generating
stronger radial flow. The same final flow can then be reached with
later starting times, giving the system more time for thermalization.
We find \cite{Song:2011hk} that the shapes of the measured pion and
proton $p_T$-spectra are well reproduced with the following parameter pairs
$(\eta/s,\,c\tau_0)$: (0,\,0.4\,fm), (0.08,\,0.6\,fm), (0.16,\,0.9\,fm),
and (0.24,\,1.2\,fm).

%
\begin{figure*}[t]
\hspace*{-3mm}
\includegraphics[height=6cm,clip=,angle=0]{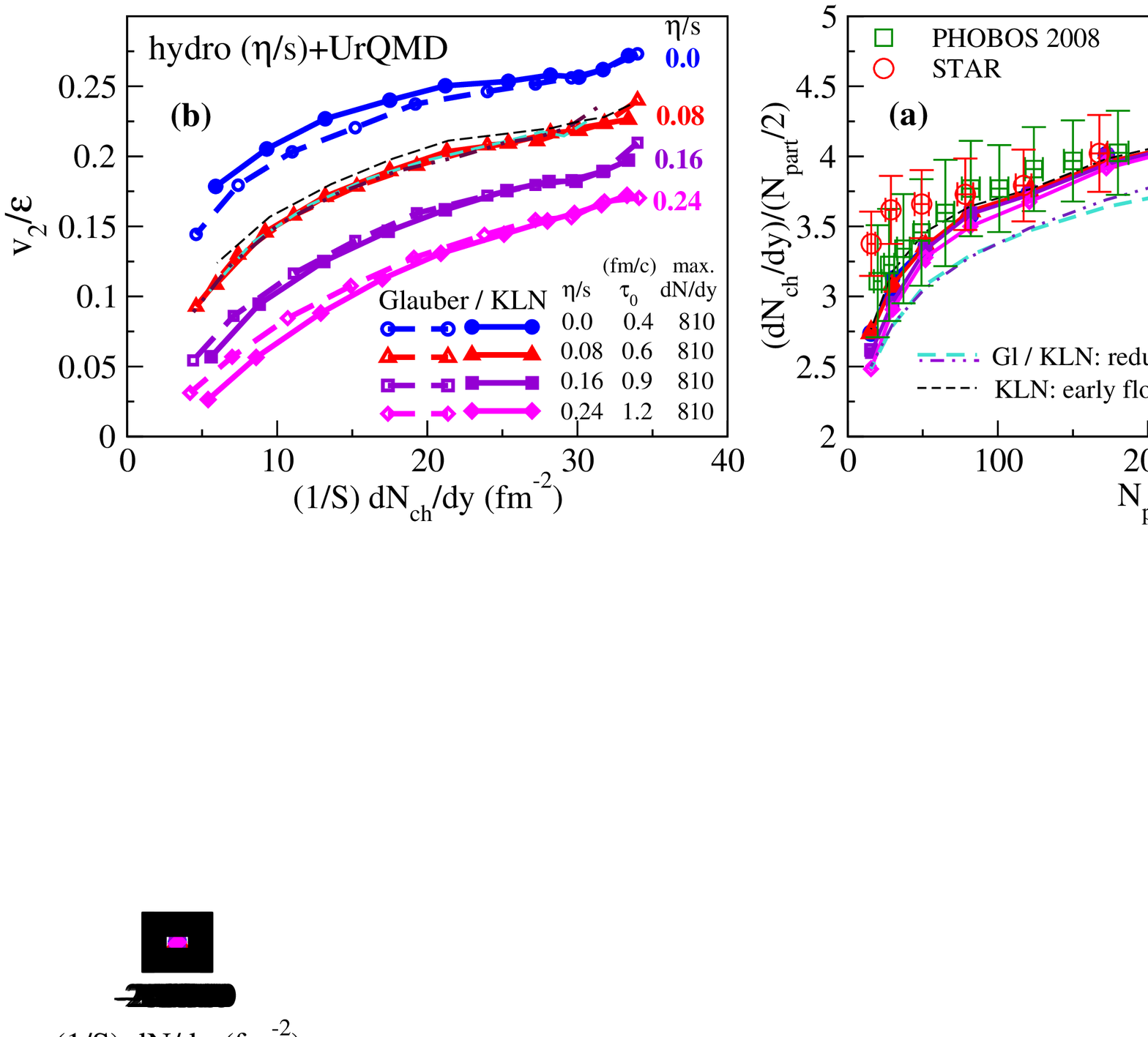}
\hspace*{-2mm}
\includegraphics[height=6cm,clip=,angle=0]{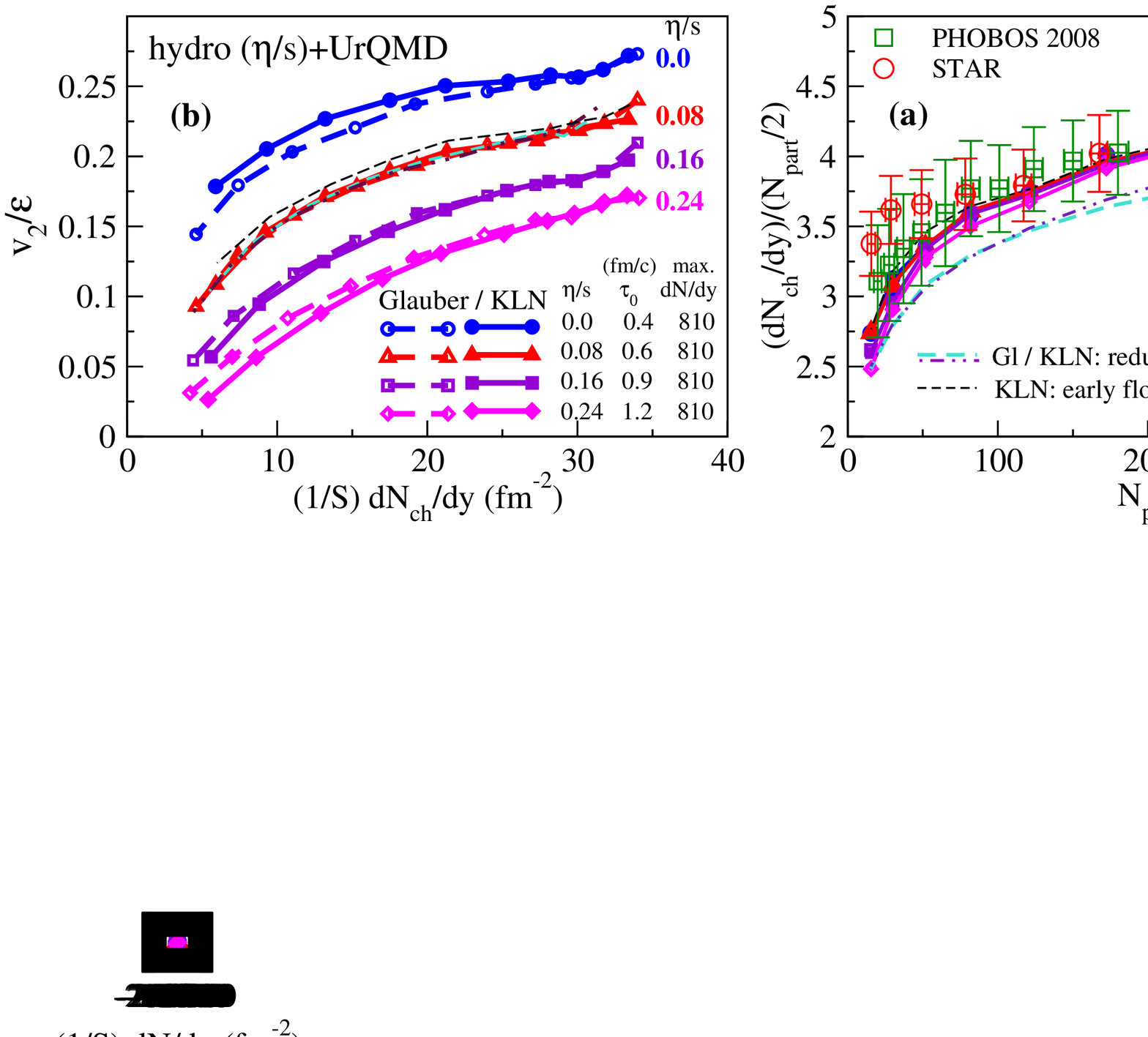}
\caption{\label{F1} (Color online)
(a) Centrality dependence of the charged hadron rapidity density per
participant pair $(\dNdy)/(N_\mathrm{part}/2)$. Experimental data are
from STAR \cite{:2008ez} and PHOBOS \cite{Back:2004dy},
using $dN_{ch}/dy\eq1.16\,dN_{ch}/d\eta$ for PHOBOS. Theoretical
lines are explained in the text. (b) Eccentricity-scaled elliptic flow
$v_2/\varepsilon$ as function of multiplicity density $(1/S)\dNdy$, for
different values of $(\eta/s)_\mathrm{QGP}$. Here and in Fig.~\ref{F2}
$v_2$ is integrated with the same cuts as in the STAR data \cite{Adams:2004bi}:
0.15\,GeV/$c{\,<\,}p_T{\,<\,}2$\,GeV/$c$, $|\eta|{\,<\,}1$. The overlap
area $S$ is always from the same initial state model as the eccentricity
$\varepsilon$ (see text). Note the universality of this theoretical relation,
independent of the model used for calculating $\varepsilon$ and $S$. Panels
(a) and (b) use the same colors and symbols but for clarity not
all corresponding curves are shown in both panels.
}
\end{figure*}
%

For each choice of $\tau_0$, the initial energy density is renormalized
to yield the same final charged hadron multiplicity $\dNdy$ in central Au+Au
collisions. Its distribution in the transverse plane is determined (via the
EOS) from the initial entropy density distribution $s(\bm{r},\tau_0;\bm{b})$
which we compute, alternatively, from two geo\-metric models discussed below.
For the shear viscous pressure tensor we use Navier-Stokes initial conditions
\cite{Song:2007fn}, noting that the system loses memory after a few
relaxation times $\tau_\pi$ where $\tau_\pi\eq\frac{3\eta}{sT} =
{\cal O}(0.2\,\mathrm{fm}/c)$ \cite{fn1b}. We ignore bulk viscosity due to
its small effect on $p_T$-spectra and $v_2$ \cite{Song:2009rh}.

The key driver for the elliptic flow generated in the collision is the
initial source eccentricity $\varepsilon\eq\frac{\La y^2{-}x^2 \Ra}
{\La y^2{+}x^2 \Ra}$ where $x$ and $y$ label the coordinates along the
short and long major axes of the fireball in the transverse plane.
$\varepsilon$ is computed from the initial entropy density after
thermalization \cite{Hirano:2009ah}. For a quantitative comparison with
experiment we account for event-by-event fluctuations of $\varepsilon$
\cite{Miller:2003kd} as follows: For each impact parameter, we
generate an ensemble of initial entropy density distributions
by Monte Carlo (MC) sampling an analytic model of the collision
geometry, recentering and rotating each distribution such that
its short major axis $x$ aligns with the direction of the
impact parameter. The plane defined by the short major axis and the beam
direction ($x{-}z$ plane) is called ``participant plane'', and the
eccentricity using this definition of $x$ is denoted as
$\varepsilon_\mathrm{part}$. Superimposing many such events yields a
relatively smooth input distribution for hydrodynamic evolution,
with an average eccentricity $\La \varepsilon_\mathrm{part} \Ra$. The
resulting elliptic flow is interpreted as the event-average $\La v_2\Ra$
for the selected centrality class.

Experimental methods for extracting the elliptic flow \cite{Voloshin:2008dg}
typically do not yield $\La v_2\Ra$. For example, the 2-particle cumulant,
denoted by $v_2\{2\}$, includes event-by-event flow fluctuations, plus
so-called ``non-flow'' contributions that are outside the purview of
hydrodynamics \cite{Alver:2008zza,Ollitrault:2009ie}. Fortunately, recent
work \cite{Ollitrault:2009ie} removed these fluctuation and
non-flow contributions from the measured elliptic flow, thereby providing
experimental values for $\La v_2\Ra$ that can be normalized by $\La
\varepsilon_\mathrm{part}\Ra$ for a direct comparison with theory.
In the absence
of non-flow, $v_2\{2\}{\,\approx\,}\sqrt{\La v_2^2\Ra}$; assuming
\cite{Ollitrault:2009ie} $\sqrt{\La v_2^2\Ra}{\,\approx\,}\frac{\La v_2\Ra}
{\La \varepsilon_\mathrm{part}\Ra}{\sqrt{\La \varepsilon_\mathrm{part}^2\Ra}}$
\cite{fn}, the experimentally determined left side can then again be
compared with the theoretically computed right side. We show such
a comparison below to check consistency.

%
\begin{figure*}[th]
\includegraphics[width=0.9\linewidth,clip=,angle=0]{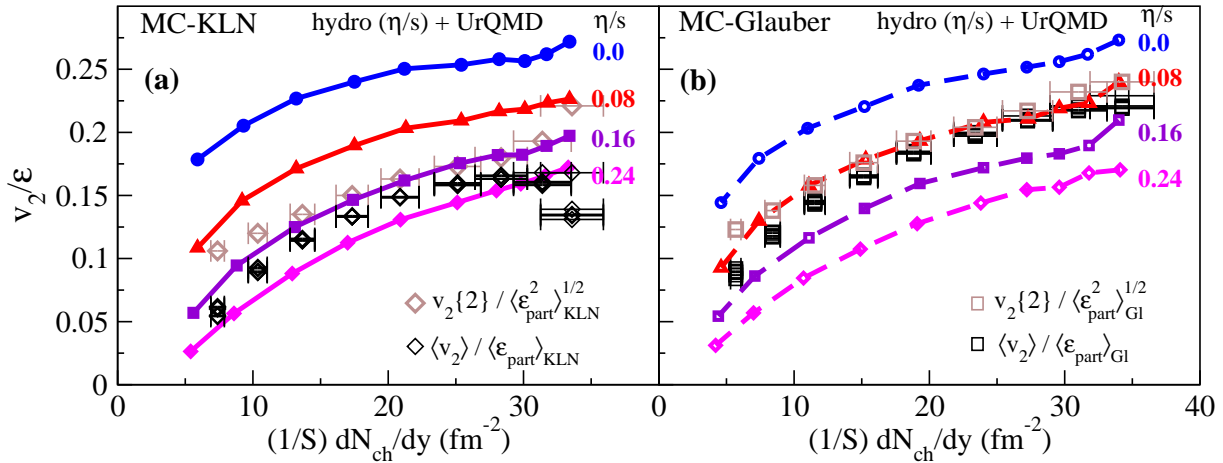}
\caption{\label{F2} (Color online) Comparison of the universal
$v_2(\eta/s)/\varepsilon$ vs. $(1/S)(\dNdy)$ curves from Fig.~\ref{F1}(b)
with experimental data for $\langle v_2\rangle$
\protect\cite{Ollitrault:2009ie}, $v_2\{2\}$ \protect\cite{Adams:2004bi},
and $\dNdy$ \protect\cite{:2008ez} from the STAR Collaboration.
The experimental data used in (a) and (b) are identical, but the
normalisation factors $\La \varepsilon_\mathrm{part}\Ra$ and $S$ used
on the vertical and horizontal axes, as well as the factor
$\La \varepsilon_\mathrm{part}^2 \Ra^{1/2}$ used to normalize
the $v_2\{2\}$ data, are taken from the MC-KLN model in (a) and from
the MC-Glauber model in (b). Theoretical curves are from simulations
with MC-KLN initial conditions in (a) and with MC-Glauber initial
conditions in (b).}
\end{figure*}
%

To compute the initial entropy density distribution in the transverse plane
we use MC versions of the Glauber \cite{Miller:2007ri} and fKLN
\cite{Drescher:2006ca} models; for a detailed description of our
procedure see \cite{Hirano:2009ah}. The models are tuned to reproduce the
measured collision centrality dependence of $\dNdy$. Figure~\ref{F1}(a) shows
that, for all permissible combinations of $\tau_0$ and $\eta/s$ and both
MC-Glauber  and MC-KLN models for the initial density distribution, the
measured centrality dependence of $\dNdy$ is well reproduced. The same
holds for the slopes of pion and proton spectra at all centralities
\cite{Song:2011hk}. (Following STAR \cite{:2008ez}, $\dNdy$ does not include
charged hyperons and weak decay products.) Two additional curves for
initial MC-Glauber and MC-KLN densities with uniformly reduced (by
$\sim10\%$) final multiplicities are shown to demonstrate that, as long
as the overall trend is preserved, small differences in $\dNdy$ extracted
from STAR, PHOBOS and PHENIX measurements do not influence our conclusions.

Figure~\ref{F1}(b) shows the key theoretical result of the present study:
the relation between eccentricity scaled elliptic flow $v_2/\varepsilon$
and multiplicity density $(1/S)\dNdy$ is approximately universal
(at least for fixed $\sqrt{s}$), depending only on the value of
$\eta/s$ for the QGP but not on any details of the model from which
$\varepsilon$ and $S\eq\pi\sqrt{\La x^2\Ra\La y^2\Ra}$ are computed. To
good approximation, switching between initial state models shifts points
for a given collision centrality along these universal curves, but not off
the lines. For example, reducing the final multiplicity by renormalizing
the initial entropy density shifts the points towards the left but also
downward because less elliptic flow is created, due to earlier hadronization.
The significantly larger $\La\varepsilon_\mathrm{part}\Ra$ from the KLN
model generates more $v_2$ than for the Glauber model, but the ratio
$v_2/\varepsilon$ is almost unchanged. Slightly larger overlap areas $S$
for the KLN sources decrease $(1/S)(\dNdy)$, but this also decreases the
initial entropy density and thus the QGP lifetime, reducing the ratio
$v_2/\varepsilon$; the result is a simultaneous shift left and downward.
Early flow \cite{fn1a} ($\tau_0\eq0.4$\,fm/$c$ for $\eta/s\eq0.08$)
increases $v_2/\varepsilon$ by $\sim5\%$, but the separation between
curves corresponding to $\eta/s$ differing by integer multiples of
$1/(4\pi)$ is much larger. Only in very peripheral collisions is
the universality of $v_2/\varepsilon$ vs. $(1/S)(\dNdy)$ slightly
broken \cite{fnA}.

The clear separation and approximate model-independence of the curves in
Fig.~\ref{F1}(b) corresponding to different $(\eta/s)_\mathrm{QGP}$ values
suggests that one should be able to extract this parameter from experimental
data. However, only $v_2$ and $\dNdy$
are experimentally measured whereas the normalization factors $\varepsilon$
and $S$ must be taken from a model. Figure~\ref{F2} shows a comparison
of the theoretical curves from Fig.~\ref{F1}(b) with STAR data normalized
by eccentricities and overlap areas taken from different initial state
models that were all tuned to correctly reproduce the centrality dependence
of $\dNdy$ shown in Fig.~\ref{F1}(a) \cite{fn2}. Since, for the same model,
the eccentricities and overlap areas depend somewhat on whether they are
calculated from the initial energy or entropy density, the same definitions
must be used in theory and when normalizing the experimental data.

Both panels of Fig.~\ref{F2} show the same data, in panel
(a) normalized by $\varepsilon$, $S$ from the MC-KLN model and in (b)
with the corresponding values from the MC-Glauber model. The theoretical
curves are from the same models as used to normalize the data.
The figure shows that comparing apples to apples matters: When comparing
the data for $v_2\{2\}/\La \varepsilon_\mathrm{part}^2\Ra^{1/2}$
with those for $\La v_2 \Ra/\La\varepsilon_\mathrm{part}\Ra$, the former
are seen to lie above the latter, showing that non-flow contributions
(which cannot be simulated hydrodynamically) either make a significant
contribution to $v_2\{2\}$ or were overcorrected in $\La v_2\Ra$
\cite{Ollitrault:2009ie}, especially in peripheral collisions. The
extraction of $\eta/s$ from a comparison with hydrodynamics thus
requires careful treatment of both fluctuation and non-flow effects.

The main insight provided by Fig.~\ref{F2} is that the theoretical curves
successfully describe the measured centrality dependence of $v_2/\varepsilon$,
i.e. its slope as a function of $\dNdy$, irrespective of whether the measured
elliptic flow is generated by an initial MC-KLN or MC-Glauber distribution.
To the best of our knowledge, the hybrid model used here to describe the
dynamical evolution of the collision fireball is the first model to achieve
this. The magnitude of the source eccentricity (and, to a lesser extent, of
the overlap area) disagrees between these two models, and this is the main
source of uncertainty for the value for $(\eta/s)_\mathrm{QGP}$ extracted
from Fig.~\ref{F2}. Both the Glauber and KLN models come in different
flavors, depending on whether the models are used to generate the initial
entropy or energy density. We have checked that the versions studied here
produce the largest difference in source eccentricity between the models.
In this sense we are confident that Figs.~\ref{F2}(a) and (b) span
the realistic range of model uncertainties for $\varepsilon$ and $S$.

We conclude that the QGP shear viscosity for
$T_\mathrm{c}{\,<\,}T{\,\alt\,}2T_\mathrm{c}$ lies within the range
$1<4\pi(\eta/s)_\mathrm{QGP}<2.5$, with the remaining uncertainty
dominated by insufficient theoretical control over the initial
source eccentricity $\varepsilon$. While this range roughly agrees
with the one extracted in \cite{Romatschke:2007mq}, the
width of the uncertainty band has been solidified by using a
more sophisticated dynamical evolution model which eliminates most
possible sources of error that the earlier analysis
\cite{Romatschke:2007mq} was unable to address. Small bulk viscous
effects \cite{Song:2009rh} and proper event-by-event hydrodynamical
evolution of fluctuating initial conditions \cite{Andrade:2008xh} may
slightly reduce the ideal fluid dynamical baseline, while pre-equilibrium
flow may slightly increase it. Although this should be studied in
more quantitative detail, we expect the quoted uncertainy band for
$(\eta/s)_\mathrm{QGP}$ to shift, after cancellations, by only a few percent.

\acknowledgments{We gratefully acknowledge fruitful discussion with
P. Huovinen, A. Poskanzer, S. Voloshin, and A. Tang. This work was
supported by the U.S.\ Department of Energy under Grants No.
DE-AC02-05CH11231, DE-FG02-05ER41367, \rm{DE-SC0004286}, and (within the
framework of the Jet Collaboration) \rm{DE-SC0004104}. T.H. acknowledges
support through Grant-in-Aid for Scientific Research No. 22740151 and
through the Excellent Young Researchers Oversea Visit Program (No. 213383)
of the Japan Society for the Promotion of Science.}



\begin{thebibliography}{99}
\bibitem{fn-1}
  We here consider only the mid-rapidity region where the approximation of
  longitudinal boost-invariance is justified. The physics at forward
  rapidity requires additional study.
\bibitem{fn2}
  We found this tuning to be an essential ingredient for a meaningful
  comparison between data and theory.
\end{thebibliography}
\end{document}